\documentstyle[aps,epsf]{revtex}
\newcommand{\bref}[1]{(\ref{#1})}
\newcommand{\vth}{\vartheta} 
\newcommand{\bea}{\begin{eqnarray}}
\newcommand{\eea}{\end{eqnarray}}
\title{Bifurcation Phenomena in Optimal Velocity Model for Traffic
 Flows}
\author{Yuji Igarashi,~~Katsumi Itoh}
\address{Faculty of Education, Niigata University, Niigata 950-2181, Japan}
\author{Ken Nakanishi}
\address{Department of Physics, Nagoya University, Nagoya 464-0814, Japan}
\author{Kazuhiro Ogura}
\address{Hosei University Daini High Schools, Kawasaki 211-0031, Japan}
\author{Ken Yokokawa}
\address{Faculty of Education, Niigata University, Niigata 950-2181, Japan}

\date{\today}

\begin{document}
\twocolumn[\hsize\textwidth\columnwidth\hsize\csname
@twocolumnfalse\endcsname
\maketitle
\begin{abstract}

In the optimal velocity model with a time lag, we show that there appear
 multiple exact solutions in some ranges of car density, describing a
 uniform flow, a stable and an unstable congested flows. This
 establishes the presence of subcritical Hopf bifurcations. Our analytic
 results have far-reaching implications for traffic flows such as
 hysteresis phenomena associated with discontinuous transitions between
 uniform and congested flows.

\end{abstract}
\hspace*{1.8cm}

\vskip1.9pc]
\narrowtext  



The optimal velocity (OV) model\cite{bando} for traffic flows attracts
considerable interest recently, since it describes congested flows as
well as uniform flows. For low mean densities, the uniform flow is
realized, while it becomes unstable above some critical density and the
congested flow shows up, with high and low density regions appearing
alternately.  At the critical density, the model predicts the presence
of a discontinuity in the correlation diagram between car density and
traffic flux, called the {\it fundamental diagram}.  This discontinuous
behavior is consistent with observed data. (In Fig.~9 of \cite{bando3}
their results are compared to an observed data\cite{JHPC}.)

Although it is an important consequence of the OV
model, the precise origin of the discontinuity has not
been identified. The purpose of this letter is to demonstrate that the
subcritical Hopf bifurcation phenomena\cite{Benjamin} is the dynamical origin
of the discontinuous transitions between uniform and congested flows.
We consider in this letter an OV model with a time lag whose exact
solutions have been found recently by three of the present
authors\cite{Igarashi}(see also ref.\cite{hasebe}). We observe that in
some ranges of density three exact solutions coexist; a stable uniform
flow, a stable and an unstable congested flows. In a numerical
simulation one of them is realized depending on its initial condition.
Any congested flow, where all cars have the same maximum ($v_{\rm max}$)
as well as the minimum value of velocities ($v_{\rm min}$), may be
characterized by the amplitude in velocity, $\Delta v =v_{\rm
max}-v_{\rm min}$.  As for the coexisting solutions, $\Delta v$ of the
stable congested flow is larger than that of the unstable flow, while
$\Delta v =0$ for the uniform flow. These values are functions of mean
headway $h$ and can be used to establish the bifurcation to be of
subcritical type.  Further the discontinuity in the traffic flux
mentioned earlier is interpreted as the hysteresis phenomenon associated
with the subcritical bifurcation.  The coexistence of stable uniform and
congested flows\cite{bando3} and the hysteresis phenomena\cite{Ogura}
were noticed earlier in simulations.  However let us emphasize here that
the subcritical bifurcation is the key concept to explain properties of
the transition between the two flows.

The system of first-order differential-difference
equations\cite{hasebebando}\cite{Igarashi} we consider is given by 
\begin{equation} 
{\dot x}_{n}(t~+~\tau)=V[\Delta x_{n}(t)]= \xi~+~\eta
\tanh\left[\left(\frac {\Delta x_{n}(t)-\rho}{2\sigma}\right)\right],
\label{DDEOM} 
\end{equation} 
where $x_n$ and $\Delta x_n =x_{n-1}-x_{n}$ correspond to the position
of $n$th car and its headway, the distance between the car and the
preceding $(n-1)$th car. $\tau$ is the time lag 
to reach the optimal velocity $V(\Delta x)$ when the traffic flow
changes.  We impose the periodic boundary conditions,
$x_{n+N}=x_{n}-L$, 
where $N$ is the total number of cars in a circuit of
length $L$. The OV function  $V(\Delta x)$ is described by a hyperbolic
tangent with four parameters, $\xi$, $\eta$, $\rho$ and $\sigma$.

The trivial solution of the system \bref{DDEOM} corresponding to a
uniform flow is given by
\begin{equation}
x^{(0)}_{n}(t)= V(h)~t - n~h,
\label{uniform}
\end{equation}
where $h=L/N$ is the common headway. A linear stability analysis shows
that the uniform flow is unstable for the region $\mid h-\rho \mid <
H_{0}$. The critical values $\rho \pm H_{0}$ are determined
by\cite{Whitham}
\begin{equation}
\frac{ \tau}{\tau_{c}} \frac{\sin\pi/N}{\pi/N}= \cosh^2\left[\frac{H_{0}}{2\sigma}\right],
\label{linear}
\end{equation}
where $\tau_{c}=\sigma/\eta$.  Thus, for $-H_{0} + \rho < h < H_{0} + \rho$,
an initially prepared uniform flow becomes unstable, and a congested
flow develops. As shown in ref.\cite{Igarashi}, the
resulting congested flow is described by an analytic function of the form
\begin{equation}
x_n(t)=  Ct~-n h~+A~{\ln} \frac
{\vth_0\left(\nu t- (2n +1)\beta+\delta, q\right)}
{\vth_0\left(\nu t- (2n +1)\beta-\delta, q\right)},
\label{xn}
\end{equation}
where $A, \beta, \nu, C, \delta$ are constants. $\vth_0(v,~q)$ is one of
the theta functions with the modulus parameter $q$. It follows from the
periodicity that $2 \beta = n_{b}/N$, where an integer $n_{b}$ is the
number of low density regions in the circuit. The width parameter $0<
2\delta <1$ determines the ratio of the length of a low density region
to $L/n_{b}$: For each length $L/n_{b}$, the length of the low density
region is given by $2\delta L/n_{b}$, while that of the high density
region by $(1-2\delta) L/n_{b}$.  In this sense, by replacing $2 \delta$
by $1-2\delta$, we exchange the length of two regions.

A crucial point for observing bifurcation is the recognition of rich
structure of the exact solution \bref{xn}. The stable uniform flows
always appear in the region $H_{0} <\mid h-\rho \mid $. Unlike this,
there must be critical values of headway $\rho \pm H_{1}$ 
($H_{1} >H_{0}$) such that the periodic solutions do not exist for 
$\mid h-\rho \mid > H_{1}$: 
no congested flows are generated if the mean density is
too high or too low. Thus, $h$ can be divided into five regions: $ h >
H_{1} +\rho $~(I), $ H_{0} +\rho <h < H_{1} +\rho $~(II), $ - H_{0}
+\rho < h < H_{0} +\rho$~(III), $-H_{1} +\rho < h < -H_{0}
+\rho$~(IV) and $ - H_{1} +\rho >h $ ~(V). We have uniform flow
solutions over the whole region; They are stable except in the region
(III).  There are no congested flow solutions in (I) and (V);  The
stable ones are known to appear in (III).  Since the stability of the
uniform flows in (II) and (IV) is confirmed against small perturbations
only, one cannot exclude the presence of other stable solutions than the
uniform flows.  We shall show that stable as well as unstable periodic
solutions do coexist actually in these regions, and that both are
described by \bref{xn}.

To this end, we discuss the relations of the model parameters $(\tau,
\xi, \eta, \rho, \sigma, N, L )$ in \bref{DDEOM} to those in the ansatz
\bref{xn}, $(A, \nu, \beta, \delta, q, C, h)$.  In addition to $L=Nh$
the following\cite{Nakanishi} should obey in order for \bref{xn} to solve \bref{DDEOM};
\begin{eqnarray}
h&-&\rho=A\ln\frac{\vth_1(2\delta-\beta, q)}{\vth_1(2\delta+\beta, q)},
\label{rho}\\
\eta&=&\frac{A\beta}{2\tau}\frac{d}{d\beta}\ln
\frac{\vth_1^2(\beta, q)}{\vth_1(2\delta+\beta, q)\vth_1(2\delta-\beta, q)},
\label{eta}\\
C&-&\xi=-\frac{A\beta}{2\tau}\frac{d}{d\beta}
\ln\frac{\vth_1(2\delta+\beta, q)}{\vth_1(2\delta-\beta, q)},
\label{xi}\\
\sigma&=&A,
\label{sigma}\\
\beta &=& \tau~\nu.
\label{Whitham}
\end{eqnarray}
Inclusion of the relation $2\beta = n_{b}/N$ with a given $n_{b}$ makes
seven reltaions between the above two sets of seven
parameters\cite{Nakanishi}: thus, we may construct exact solutions for a
given set of the model parameters.  In this paper we restrict ourselves
to the case $n_{b}=1$, for simplicity. (See ref.\cite{Nakanishi} for
solutions with $n_{b}\ne 1$.)  The eq.\bref{Whitham} is the Whitham's
dispersion relation\cite{Whitham}, an important characteristic of the
delayed model.  Note that $h-\rho$ and $C-\xi$ change their signs while
$\eta$ remains invariant when we replace $2 \delta$ by $1-2\delta$.

We show the presence of multiple solutions for a given set, $(\tau,~
\xi,~\eta,~\rho,~\sigma,~N,~L)$. To obtain the parameters of the
solutions, one first fixes $(A,~\nu,~\beta)$ from \bref{sigma},
\bref{Whitham} and $2\beta N =1$.  As the modulus parameter, we use
$\kappa=- \pi/(\ln q)$ instead of $q$.  By solving \bref{eta}, $\delta$
is computed as a function of $\kappa$. (As a result of the property of
$\delta$ discussed earlier, it has two branches, $\delta_{1}(\kappa)$
and $\delta_{2}(\kappa)$ such that $2\delta_{1}(\kappa)= 1-
2\delta_{2}(\kappa)$.) Then substituting $\delta(\kappa)$ to \bref{rho}
gives us a function, $h(\kappa)$. The result is shown in Fig.1.

\begin{figure}
\leavevmode
\epsfxsize=8cm
\setlength{\unitlength}{0.76cm}
\begin{picture}(10.5,10)
\put(0.5,0){\epsfbox{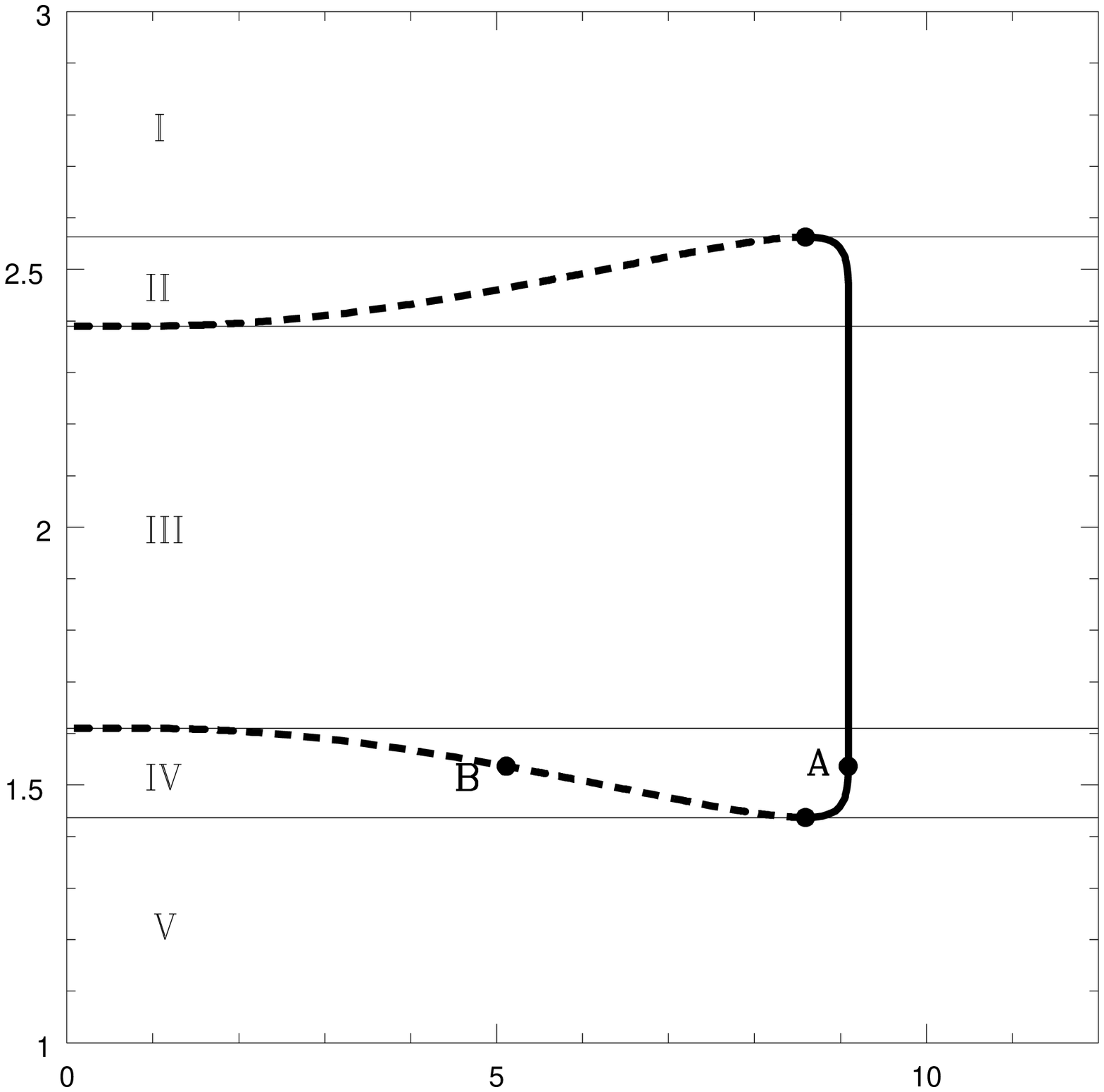}} 
\put(5.4,0.15){$\kappa$}
\put(0.6,5.39){$\it h$}
\put(0.1,7.15){\footnotesize $\it \rho+H_{0}$}
\put(0.1,7.95){\footnotesize $\it \rho+H_{1}$}
\end{picture}
\caption{The relation between the mean headway $h$ and $\kappa$.  The solid
and broken lines correspond to stable and unstable solutions,
respectively.  Here the parameters are $\tau=0.58228, ~
\xi=\tanh 2, ~\eta=1, ~ \rho=2, ~ \sigma =1/2$ and $N=20$,
for which $H_{0}=0.38978$ and $H_{1}=0.56290$.}
\end{figure}

As shown in Fig.1, its graph is symmetric around the axis $h=\rho$.
At the symmetric point, $2\delta$ becomes 1/2. It should be noted that
$\kappa(h)$ becomes a two-valued function in the regions
(II) and (IV). In each region, we have two periodic solutions for a
given set of the model parameters. We have confirmed by numerical
simulations that one (denoted by A on the solid line in Fig.1) is
stable, while the other (denoted by B on the broken line in Fig.1) is
unstable: When a small perturbation is added, the latter goes down to
the former or to a uniform flow with the same value of $h$.
Therefore, there are altogether three solutions in this region,
indicating a subcritical bifurcation.  The bifurcation points $h= \rho
\pm H_{0}$ are determined from \bref{rho} and \bref{eta}
for $\kappa =0$, which end up with \bref{linear}.  The value of
$\kappa$ which gives the critical values $h= \rho \pm H_{1}$ is
determined by
\begin{eqnarray}
[Z(B)&+&\frac{{\rm cn}(2B)+ {\rm dn}(2B)-1}{{\rm sn}(2B)}
-\frac{Z'(B)}{2Z(B)}]  \\
\times [ \frac{\tau}{B\tau_c}&-&
\frac{{\rm cn}(2B)+{\rm dn}(2B)}{{\rm sn}(2B)}] 
 = \frac{Z(2B)-Z(B)}{{\rm sn}(2B)}- \frac{Z'(B)}{2}, \nonumber
\label{crimod}
\end{eqnarray}
where $B=2K\beta$, $Z$ is the Jacobi's zeta function and $K$ is the
complete elliptic integral of the first kind.  This determines $H_{1}$.

As an ``order parameter'' characterizing the subcritical bifurcations,
 we may take the velocity amplitudes $\Delta v$ of relevant flows as
 functions of $h$.  In Fig.~2 we have drawn the analytical results
 with the solid and broken lines, which agree with numerical
 simulations shown with the small squares.  The simulations are done
 by gradually increasing $h$ and the results for $\Delta v$ change as
 indicated with thin arrows: when we increase $h$, there is a jump in
 $\Delta v$ at $h=\rho- H_{0}$, from a uniform flow to a congested
 flow; at $h=\rho+ H_{1}$ there is another jump from a congested flow
 to a uniform flow. (Although not indicated in the figure, jumps occur
 at $h=\rho+ H_{0}$ and $h=\rho-H_{1}$ for decreasing $h$.) This is
 the hysteresis phenomenon associated with the subcritical
 bifurcation.  It is appropriate to explain how we performed the
 simulations.  Suppose we have a stable flow as a result of the
 simulation at a value for the mean headway $h$, say $h_0$.  Taking a
 snapshot of the flow, we know the positions of cars at the time. Then
 we measure the headway of each car and increase it by $\Delta h$, and that
 configuration is used as the initial condition for the simulation with
 $h=h_0+\Delta h$.  This procedure gives the results for the region
 (II) staying on the curve for the congested flow: it is important to
 keep the $\Delta h$ small enough since the results would heavily
 depend on the initially prepared configurations\cite{Ogura}.

\begin{figure}
\leavevmode
\epsfxsize=8cm
\setlength{\unitlength}{0.76cm}
\begin{picture}(10.3,10.7)
\put(0.3,0){\epsfbox{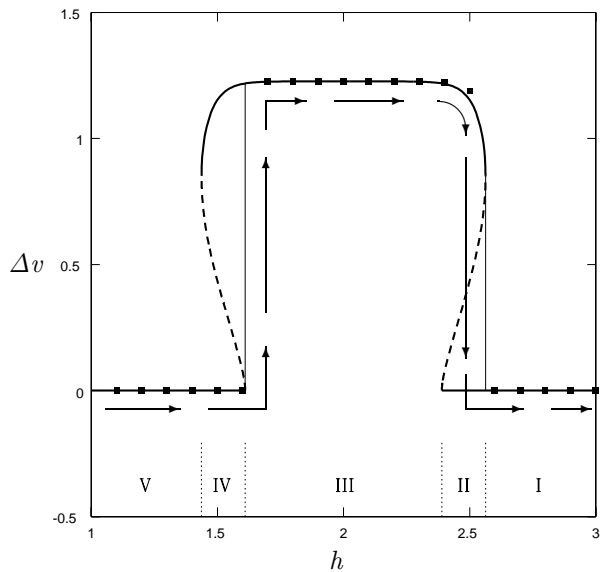}} 
\put(5.61,0.2){$\it h$}
\put(0,5.4){$\it \Delta v$}
\put(1.7,3){\vector(1,0){1.3}}
\put(3.5,3){\line(1,0){1}}
\put(4.5,3){\vector(0,1){1.1}}
\put(4.5,4.7){\vector(0,1){2.7}}
\put(4.5,7.9){\line(0,1){0.5}}
\put(4.5,8.4){\vector(1,0){0.7}}
\put(5.7,8.4){\vector(1,0){1.2}}
\put(7.5,7.9){\oval(1,1)[tr]}
\put(8,7.9){\vector(0,-1){0.1}}
\put(8,7.4){\vector(0,-1){3.5}}
\put(8,3.6){\line(0,-1){0.6}}
\put(8,3){\vector(1,0){1}}
\put(9.5,3){\vector(1,0){0.7}}
\end{picture}
\caption{The $h$-$\Delta v$ relation from the exact solution, compared
 with a numerical study.  The simulations were done by increasing $h$ as
 indicated by arrows, the results are plotted with the small squares.
 We observe a hysteresis phenomenon.}
\end{figure}

We discuss now flux-density correlation diagram, the fundamental
diagram. The flux $Q$ may be defined by the number of
cars passing through some reference point in a certain time
interval. For a uniform flow, it is given by product of the density
$1/h$ and the common velocity $V(h)$:
\bea
Q=V(h)/h.
\label{uflux}
\eea
We calculate the flux of a congested flow as follows.  Since high
density regions are moving backward with velocity $v_{B}$, it will be
convenient to take the rest frame of the density waves.  The period
$T_0$ defined for a car to make a round in the rest frame is given by
\bea
T_0 = \frac{1}{\nu} = 2\tau ~N, 
\label{T0}
\eea
where we have used the Whitham's relation, $\beta = \tau\nu = 1/(2N)$.  
Since the average velocity of cars in the original coordinate is to be
$L/T_0 - v_{B}$, the flux $Q$ may be given by
\bea
Q=\frac{1}{h} \left(\frac{L}{T_0}- v_{B}\right)=  \frac{1}{2\tau}~-~v_B/h .
\label{cflux}
\eea
Let us calculate $v_{B}$ by considering the repetitive spatio-temporal
pattern in a congested flow:
\bea
x_{n-1}(t)= x_{n}(t+2\tau)+ 2~v_{B}~\tau.
\label{xn-1}
\eea
This implies that the $x_{n}$ has the same time dependence as $x_{n-1}$
apart from a time delay $2\tau$ and a distance $2v_{B}\tau$. It follows
from \bref{xn} and \bref{xn-1} that
\bea
v_{B}= \frac{h}{2\tau}~-~C .
\label{C}
\eea
Therefore, one obtains a simple expression for $Q$
\bea
Q= C/h .  
\label{C}
\eea
As discussed previously, one calculates the parameter $C$ as a
function of $\kappa$ using \bref{xi}. Together with $h(\kappa)$, it
gives $C(h)$.  Our analytic results are shown in Fig.3:
Discontinuities in the traffic flow show up as a result of a
hysteresis phenomenon, which is consistent with the observed data
qualitatively.  Since the size of the discontinuous jumps may be
determined analytically, a quantitive comparison is also possible.

A few comments are in order. (1) The presence of the subcritical Hopf
bifurcations naturally define the boundaries between (I) and (II), and
between (IV) and (V).  This should be compared with the Fig.~8 of
\cite{bando3}.  (2) In Fig.3, the discontinuities in the low density
side appears after passing through the uniform-flow's peak.  Since the
position of the peak is determined by the OV function, we may take a set
of model parameters so that the discontinuities appear before the peak.

\begin{figure}
\leavevmode
\epsfxsize=8cm
\setlength{\unitlength}{0.6cm}
\begin{picture}(10,13)
\put(0,0){\epsfbox{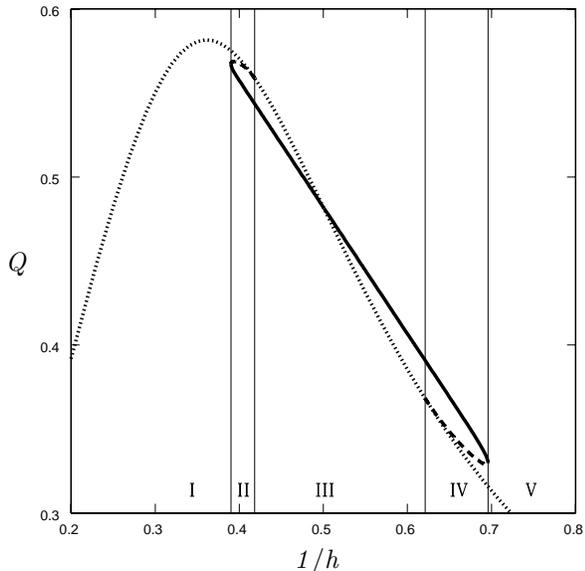}} 
\put(6.4,0.2){$\it 1/h$}
\put(0,6.8){$\it Q$}
\end{picture}
\caption{The fundamental diagram for traffic flow.  The dotted line is
 for uniform flows, while the solid and broken lines are for stable and
 unstable congested flows, respectively.}
\end{figure}

In summary, the presence of subcritical Hopf bifurcations in a delayed
 optimal velocity model which admits exact solutions is established by
 showing coexistence of stable and unstable solutions. Although our
 results are obtained in a specific OV model, we believe that any OV
 type model will show up the subcritical bifurcations. Actually, we have
 confirmed by numerical simulations the presence of multiple stable
 solutions for the model given by ${\ddot x}_{n} = a [V(\Delta x_{n})-
 {\dot x_{n}}]$(First studied in \cite{Ogura} in this
 context). Therefore, our results given here should be considered, at
 least qualitatively, to be a universal feature of the OV models
 describing spatio-temporal patterns.  They may be of fundamental
 importance in the area of the traffic engineering.  In order to confirm
 the subcritical bifurcation directly highly desirable are more precise
 data of the $h$-$\Delta v$ relation.

Parts of our numerical works were done with the computer fascilities of
Yukawa Institute for Theoretical Physics.  We are grateful to
K. Nishinari, H. Hayakawa and Y. Ono for discussions and suggestions.


\end{document}